\documentclass[3p]{elsarticle}
\usepackage[utf8]{inputenc}
\usepackage[english]{babel}
\usepackage{graphicx}
\usepackage{dcolumn}
\usepackage[table,xcdraw]{xcolor}
\usepackage{bm}
\usepackage{natbib}
\usepackage{latexsym}
\usepackage{mathrsfs}
\usepackage{amssymb}
\usepackage{amsmath}
\usepackage{amscd}
\usepackage{color}
\usepackage{pifont}
\usepackage{verbatim}
\usepackage{placeins}
\usepackage[T1]{fontenc}
\usepackage{hyperref}
\usepackage{ulem}
\usepackage{multirow}
\hypersetup{
  colorlinks   = true, %colors links instead of ugly boxes
  urlcolor     = black, %color for external hyperlinks
  linkcolor    = red, %color of internal links
  citecolor    = blue %color of citations
}
\definecolor{magenta}{rgb}{1.0, 0.0, 1.0}
\definecolor{orange}{rgb}{0.98, 0.6, 0.01}
\definecolor{brown}{rgb}{0.59, 0.29, 0.0}
\definecolor{aquamarine}{rgb}{0.5, 1.0, 0.83}
\definecolor{blue-violet}{rgb}{0.54, 0.17, 0.89}
\definecolor{verde}{rgb}{0.04, 0.6, 0.02}
\definecolor{cinza}{rgb}{0.6, 0.6, 0.6}

\usepackage{lineno,hyperref}
\modulolinenumbers[5]

\begin{document}

\begin{frontmatter}

%\title{The decisive role of containment measures against COVID-19
%  applied ahead of time}

\title{How relevant is the decision of containment measures against COVID-19
 applied ahead of time?}
\author{Eduardo L.~Brugnago$^{1}$}
\ead{elb@fisica.ufpr.br}
\author{Rafael M.~da Silva$^{1}$}
\ead{rmarques@fisica.ufpr.br}
\author{Cesar Manchein$^2$}
\ead{cesar.manchein@udesc.br}
%\author{\cbeims{Carlos F.O.~Mendes$^3$}}
%\ead{cfabio.mendes@gmail.com}
\author{Marcus W.~Beims$^{1}$}
\ead{mbeims@fisica.ufpr.br}
\address{$^1$Departamento de F\'\i sica, Universidade Federal do
  Paran\'a, 81531-980 Curitiba, PR, Brazil}
\address{$^2$Departamento de F\'\i sica, Universidade do Estado de
  Santa Catarina, 89219-710 Joinville, SC, Brazil}
%
%\address{$^3$Escola Normal Superior, Universidade do Estado do Amazonas, 
%69050-010 Manaus, AM,  Brazil}
%
\date{\today}

\begin{abstract}
The cumulative number of confirmed infected individuals by the new coronavirus outbreak 
until April 30$^{\text{th}}$, 2020, is presented for the countries: Belgium, Brazil, 
United Kingdom (UK), and United States of America (USA). After an initial period with a 
low incidence of newly infected people, a power-law growth of the number of confirmed cases 
is observed. For each country, a distinct growth exponent is obtained. For Belgium, UK, and 
USA, countries with a large number of infected people, after the power-law growth a distinct 
behavior is obtained when approaching saturation. Brazil is still in the power-law regime. 
Such updates of the data and projections corroborate recent results regarding the power-law 
growth of the virus and their strong Distance Correlation between some countries around the 
world. Furthermore, we show that act in time is one of the most relevant non-pharmacological 
weapons that the health organizations have in the battle against the COVID-19, infectious 
disease caused by the most recently discovered coronavirus. We study how changing the social 
distance and the number of daily tests to identify infected asymptomatic individuals can 
interfere in the number of confirmed cases of COVID-19 when applied in three distinct days, 
namely April 16$^{\text{th}}$ (early), April 30$^{\text{th}}$ (current), and May 
14$^{\text{th}}$ (late). Results show that containment actions are necessary to flatten the 
curves and should be applied as soon as possible.
%The social distance parameter and the test numbers vary between strong and soft social distance 
%degrees and strong and soft number of daily dentified infected individuals.
\end{abstract}


\end{frontmatter}

%%%%%%%%%%%%%%%%%%%%%%%%%%%%%%%%%%%%%%%%%%%%%%%%%%%%%%%%%%%%%%%
\section{Introduction}
\label{intro}
%%%%%%%%%%%%%%%%%%%%%%%%%%%%%%%%%%%%%%%%%%%%%%%%%%%%%%%%%%%%%%%

Since the first infection of the coronavirus in December 2019, observed in Wuhan (China), 
the virus has spread around the world very quickly and nowadays 215 countries, areas, or 
territories\footnote{``Territories'' include territories, areas, overseas dependencies and 
other jurisdictions of similar status \cite{OMS}.} report confirmed cases of the infection. 
Innumerable scientists in distinct areas are using their knowledge in the battle against 
the still evolving COVID-19 outbreak around the globe. The daily analysis of data about 
the spreading of the virus and possible interpretations that allow us to track and control 
the virus are of most relevance. It is a timely appeal to find explanations and models 
which may allow us to better understand the evolution of the viruses, saving lives, and 
avoiding economic and social catastrophes \cite{puevo20}.

In the battle against the COVID-19 spreading, some models focus on the geographical
spread of the virus \cite{blasius20,arenas20}, while others remain restricted to a given 
area, or country, but analyze the local temporal development of the epidemic. In the 
context of diseases, in 1760 Daniel Bernoulli proposed a mathematical model of disease 
propagation and showed the efficiency of the preventive inoculation technique against 
smallpox \cite{ber1760}. This model included susceptible and immune individuals 
\cite{dietz02}. Later on, Kermack and McKendrick \cite{ker1927} came up with the 
Susceptible-Infected-Recovered (SIR) model. During the last years, other more sophisticated 
models have been proposed like the Susceptible-Exposed-Infected-Recovered (SEIR) model 
\cite{modeloSeiqr1,modeloSeiqr2,modeloSARS} and its modified versions 
\cite{elb20-1,modeloSirx,modeloShir,modeloSiqr,modeloSeir2}. Both approaches, the 
geographical spread, and the local temporal evolution are of most relevance.

It is well-known that the decisive quantity used to regulate the dynamical evolution of 
epidemics, in general, is the average reproductive number $R_0$, which gives the number of 
secondary infected individuals generated by a primary infected individual. While for 
values $R_0 < 1$ the number of newly infected individuals decreases exponentially, for 
$1 < R_0 < \infty$ it increases exponentially \cite{vas06, Rfactor}. Starting from the 
primordial exponential solution put up by Verhulst in 1838, the well known logistic model 
for the {\it law of population growth} \cite{baca11}, models were improved more and more 
in the last decades to better describe the nonlinear and complex comportments which occur 
in our environment. In fact, in many realistic systems, power-law functions are the {\it 
law of growth (or decrease)}, as in the branching processes with a diverging reproductive 
number \cite{vas06}, in scale free networks and small worlds \cite{watts04}, and in 
foraging in biological systems \cite{gosma11}. Indeed, recent investigations showed a 
power-law growth of the cumulative number of infected individuals by the new coronavirus 
\cite{elb20-1, modeloSirx,singer20-2, Marsland2020}, which might be typical of small world 
networks \cite{ray20} and possibly related to fractal kinetics and graph theory 
\cite{ziff20}. 

Recently, we have shown that power-law growth is observed in countries from four distinct 
continents \cite{elb20-1} until March 27$^{\text{th}}$, 2020. The considered countries 
were: Brazil, China, Germany, Italy, France, Japan, Spain, Republic of Korea, and the 
United States of America (USA). One leading observation was that after an initial time 
with a low incidence of newly infected people, the growth of the cumulative number of 
confirmed cases for all studied countries followed a power-law. The Distance Correlation 
\cite{Szek07,Szek13,CFMWB,carlos19} between these countries was found to be very strong 
and suggest a universal characteristic of the virus spreading. One of the goals of the 
present work is to update to April 30$^{\text{th}}$, 2020 the time-series analysis for the 
COVID-19 growth for the countries Brazil and USA. We included Belgium and United Kingdom 
(UK) on this list and leave out the other countries which are reaching the saturation 
regime.  

\begin{figure}[!t]
  \centering
  \includegraphics[width=0.85\columnwidth]{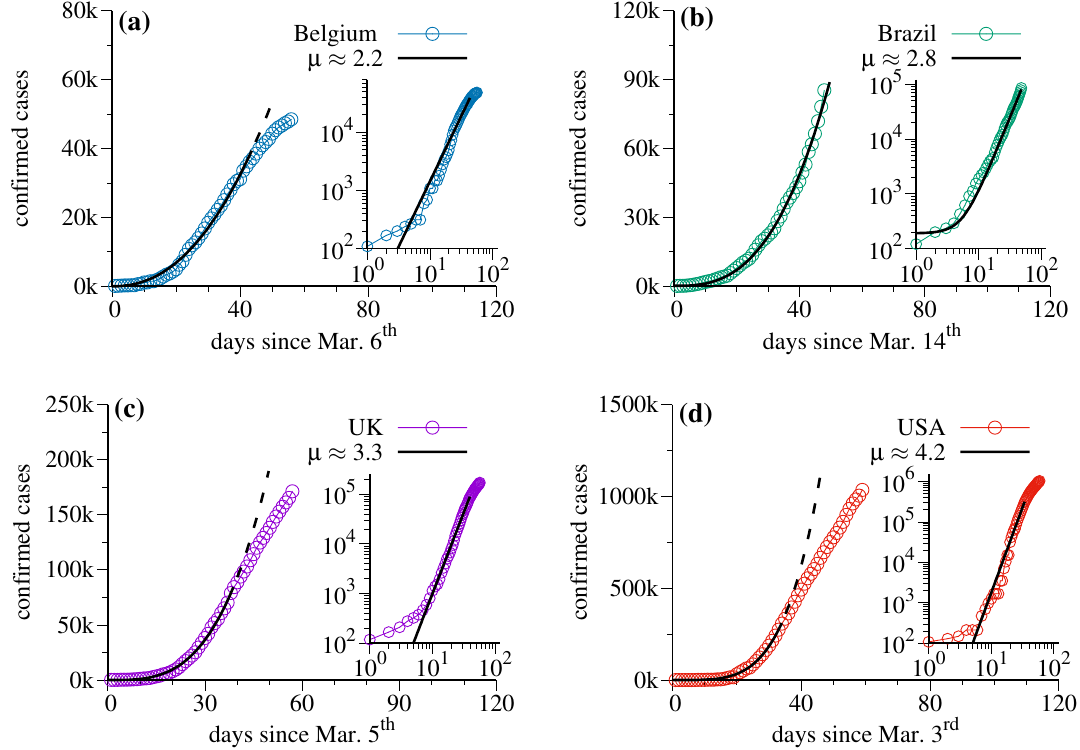}
  \caption{The cumulative number of confirmed cases of COVID-19 (empty circles) as a 
  function of time for (a) Belgium, (b) Brazil, (c) UK, and (d) USA, excluding days with 
  less than 100 infected. The black-continuous curves represent the function 
  $\propto t^{\mu}$ that fit the time-series, with exponent $\mu$ for each country. The 
      insets display the same curves but in the log-log plot.}  
    \label{rat}
  \end{figure}

Figure~\ref{rat} displays the cumulative number of confirmed cases of COVID-19 (empty 
circles) as a function of time for four exemplary countries: Belgium [Fig. \ref{rat}(a)], 
Brazil [Fig. \ref{rat}(b)], UK [Fig. \ref{rat}(c)], and USA [Fig \ref{rat}(d)]. Data were 
collected from the situation reports published daily by the World Health Organization 
(WHO)~\cite{OMS}. We call to attention that the values in the vertical axis in 
Fig.~\ref{rat} change for different countries. Initial data, regarding the days with less 
than 100 infected individuals, were discarded. The black-continuous curves represent the 
function $\propto t^{\mu}$ that fits the time series and the exponent $\mu$ for each 
country is indicated in each panel. The insets display the same curves but in the log-log 
plot. Straight lines in the log-log plot indicate power-law growth. The only country for 
which the power-law growth still takes place is Brazil, as shown in Fig. \ref{rat}(b). The 
reason is that it is still away from the saturation point. This is different for Belgium, 
UK, and USA, as can be seen in Figs.~\ref{rat}(a), \ref{rat}(c), and \ref{rat}(d), 
respectively. Dashed-black lines in these three panels are projections in case the 
power-law would have guided the growth. 

%\FloatBarrier

Besides the above updates, in this paper, we describe in detail the modified SEIR model 
which was used recently~\cite{elb20-1} to propose strategies to flatten the power-law 
curves. It is shown how to adjust the parameters of the model to real data. Furthermore, 
using the same model we discuss what would be the effect of early, current, and late 
non-pharmacological actions to flatten the curves of the four countries shown in 
Fig.~\ref{rat}. This clearly shows that each day lost by delaying non-pharmacological 
actions can cost many lives. 

The paper is divided as follows. In Sec.~\ref{model} we present in detail the model used 
in this work. Section~\ref{ppf} discusses the effect of containment actions on the total 
number of confirmed infected cases applied in three distinct days and 
Sec.~\ref{conclusions} summarizes our results.

%%%%%%%%%%%%%%%%%%%%%%%%%%%%%%%%%%%%%%%%%%%%%%%%%%%%%%%%%%%%%%%
\section{The model}
\label{model}
%%%%%%%%%%%%%%%%%%%%%%%%%%%%%%%%%%%%%%%%%%%%%%%%%%%%%%%%%%%%%%%

\subsection{Equations, variables and parameters}

In this section we describe in detail the model used to reproduce the realistic data of 
the WHO and to predict the effect of strategies used to flatten the power-law curves. The 
model that we used is the modified SEIR model described by the following six Ordinary 
Differential Equations (ODEs) \cite{elb20-1}
%\begin{align}
%    \dot{S}   &= -\frac{\theta}{T_{inf}}\frac{(I_s + \alpha I_a)}{N}S,\\
%    \dot{E}   &= \frac{\theta}{T_{inf}}\frac{(I_s + \alpha I_a)}{N}S - \frac{E}{T_{lat}},\\
%    \dot{I}_s &= (1 - \beta)\frac{E}{T_{lat}} - \left(\kappa_s + \frac{1}{T_{inf}}\right)I_s,\\
%    \dot{I}_a &= \beta\frac{E}{T_{lat}} - \left(\kappa_a +  \frac{1}{T_{inf}}\right)I_a,\\
%    \dot{Q}   &= \kappa_s I_s + \kappa_a I_a - \frac{Q}{T_{ser}},\label{quarentena}\\
%    \dot{R}   &= \frac{I_s + I_a}{T_{inf}} + \frac{Q}{T_{ser}}.
%	\dot{R}_i &= \frac{1}{T_{inf}}\left(I_s + Q \right),\\
%	\dot{R}_a &= \frac{I_a}{T_{inf}}.
%\end{align}
\begin{align}
	\frac{dS}{dt}   &= -\frac{\theta}{T_{inf}}\frac{(I_s + \alpha I_a)}{N}S,\\
	\frac{dE}{dt}   &= \frac{\theta}{T_{inf}}\frac{(I_s + \alpha I_a)}{N}S - \frac{E}{T_{lat}},\\
	\frac{dI_s}{dt} &= (1 - \beta)\frac{E}{T_{lat}} - \left(\kappa_s + \frac{1}{T_{inf}}\right)I_s,\\
	\frac{dI_a}{dt} &= \beta\frac{E}{T_{lat}} - \left(\kappa_a +  \frac{1}{T_{inf}}\right)I_a,\\
	\frac{dQ}{dt}   &= \kappa_s I_s + \kappa_a I_a - \frac{Q}{T_{ser}},\label{quarentena}\\
	\frac{dR}{dt}   &= \frac{I_s + I_a}{T_{inf}} + \frac{Q}{T_{ser}}.
\end{align}
In addition to these equations, we compute the cumulative number of confirmed cases $C$ of 
COVID-19 from the following ODE:
\begin{align}
	\frac{dC}{dt} = (1 - \beta)\frac{E}{T_{lat}} + \kappa_a I_a.
\end{align}
Through this variable, the parameters $(\theta, \kappa_s)$ can be adjusted, as 
described later on. Table \ref{tabelaVariaveis} brings together all variables and 
parameters of the model and their meaning. In the case of the variables, the initial 
conditions are also presented and in the case of the parameters, the predefined values 
obtained from preceding studies are also listed. The highlight lines in 
Table~\ref{tabelaVariaveis} call to attention to the variable $C$, which is the main 
quantity analyzed in this work, to the adjustable parameters $\theta$ and $\kappa_s$, and 
to the strategic parameter $\kappa_a$. After the adjustment, the parameters $\theta$ and 
$\kappa_a$ will be varied to give rise to specific strategies. Worth to mention 
that $\theta=\gamma R_0$, where $R_0$ is the basic reproductive number without social 
distance actions, and $\gamma$ is the interaction factor between individuals. This factor 
can be interpreted as the ratio between the average of the daily social interaction due to 
the applied social distance actions and the case with no actions at all. Larger social 
distance implies smaller values of $\theta$, which is equivalent to reduce $R_0$. The 
distinction between $\theta$ and $R_0$ allows us to identify the direct effects of the 
actions in the battle against the pandemic. Thus, the ideal situation would be to find 
$\theta<1$.

\begin{table}[!t]
	\caption{{Variables and parameters (with their meaning) of the model with initial 
	conditions and predefined values, respectively.}}
	\centering
	\resizebox{\textwidth}{!}{
	\begin{tabular}{|c|l|l|}\hline
		\rowcolor[rgb]{0.8,0.8,0.8}	
		\textbf{Variable} & \textbf{Meaning} & \textbf{Initial condition} \\ \hline
		$N$      & Country population.               & Depends on the country         \\ \hline
		$S$      & Individuals susceptible to infection.   & $S(t_0)=N-E(t_0)-C(t_0)/(1-\beta)$\\ \hline
		$E$      & Exposed individuals, latent cases.      & Adjusted from data         \\ \hline
		$I_s$    & Symptomatic infectious cases.           & $I_s(t_0)=C(t_0)$            \\ \hline
		$I_a$    & Asymptomatic and mild infectious cases.  & $I_a(t_0)=\beta C(t_0)/(1-\beta)$ \\ \hline
		$Q$      & Isolated individuals.                   & $Q(t_0)=0$                   \\ \hline
		$R$      & Recovered individuals. Became imune.    & $R(t_0)=0$                   \\ \hline
		\rowcolor[rgb]{1.0,0.7,0.7}	\hline
		$C$    & Total of confirmed cases.              & WHO data~\cite{OMS} \\ \hline \hline
		\rowcolor[rgb]{0.8,0.8,0.8}	
		\textbf{Parameter} & \textbf{Meaning} & \textbf{Class} \\ \hline
		$T_{ser}=7.5$      & Serial interval.                              & Predefined~\cite{tempoSerialIncubacao} \\ \hline
		$T_{lat}=5.2$      & Mean incubation period.                       & Predefined~\cite{modeloSeir2,tempoSerialIncubacao} \\ \hline
		$T_{inf}=2.3$      & Infectious period. $T_{inf}=T_{ser}-T_{lat}$  & Predefined~\cite{modeloSeir2} \\ \hline
		$\alpha = 1$       & Ratio between infectiousness of $I_a$ and $I_s$.  & Predefined \\ \hline
		$\beta = 0.8$      & Population ratio which remains asymptomatic.  & Predefined~\cite{OMS} \\ \hline\hline
		 \rowcolor[rgb]{0.44, 0.77, 0.99}
		$\theta=\gamma R_0$           & $R_0$ the reproduction number, $\gamma$ is the interaction factor. & Adjustable/Strategies \\ \hline \rowcolor[rgb]{0.04, 0.97, 0.02}	
		$\kappa _s$        & Isolation rate of symptomatic individuals.    & Adjustable/Strategies \\ \hline
		\rowcolor[rgb]{0.54, 0.97, 0.89}
		$\kappa _a$        & Isolation rate of asymptomatic individuals.    & Strategies \\ \hline
	\end{tabular}}
	\label{tabelaVariaveis}
\end{table}

\begin{figure}[!t]
  	\centering
  	\includegraphics[width=0.6\columnwidth]{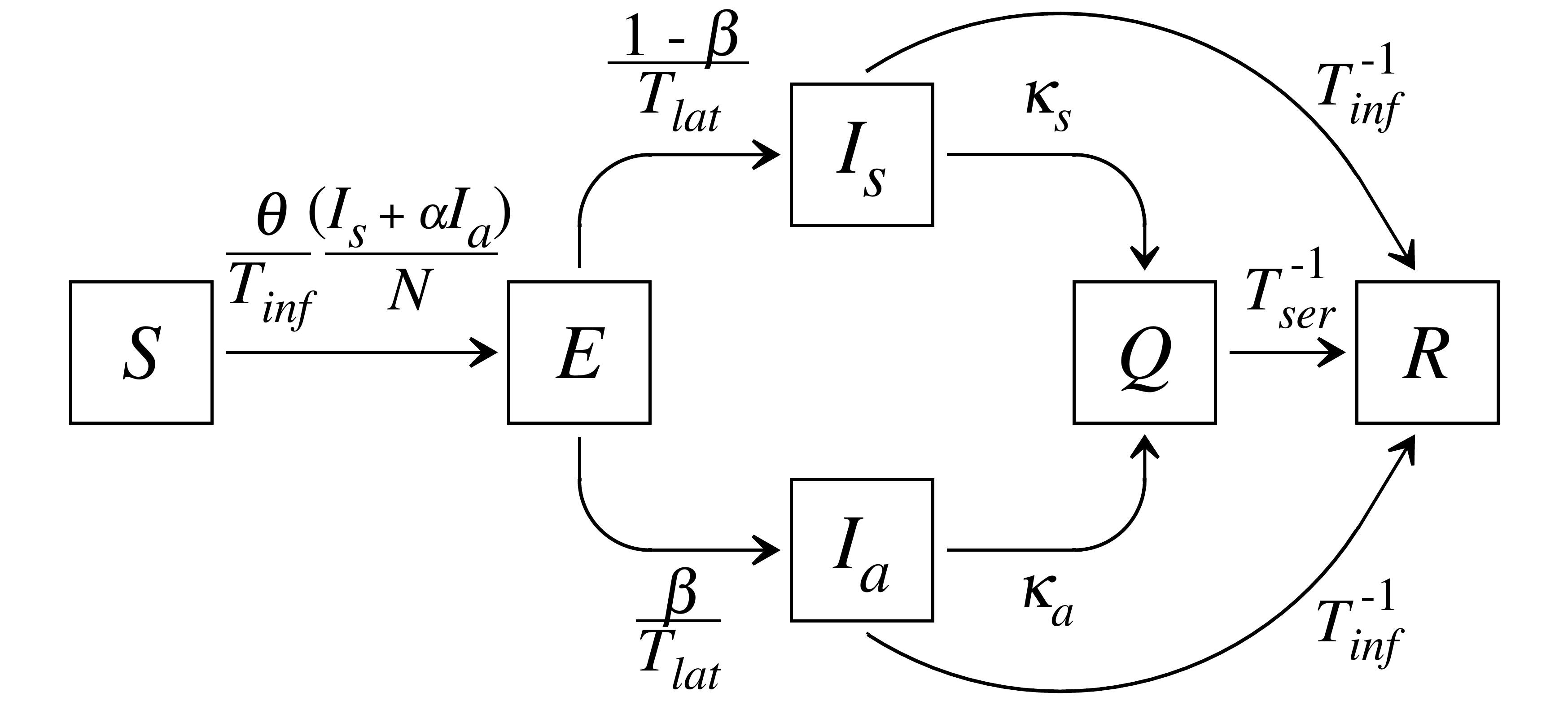}
  	\caption{Schematic representation of the modified SEIR model, their variables, and 
    the role of the parameters connecting the variables.}
  	\label{esquema}
\end{figure}

Figure~\ref{esquema} is a schematic representation of the model, showing the variables and 
the connections between them through the parameters. Condensing the explanation of the 
schema, starting from the left, susceptible individuals $S$ develop into exposed 
individuals $E$ by a rate $\theta(I_s+\alpha I_a)/(N T_{inf})$ which, after a latent time 
$T_{lat}$, become symptomatic $I_s$ or asymptomatic $I_a$ with the rate $(1-\beta)/T_{lat}$ 
and $\beta/T_{lat}$, respectively. Applying daily tests in a rate $\kappa_s$ ($\kappa_a$) 
to identify symptomatic (asymptomatic) infected individuals, they are immediately sent to 
quarantine $Q$, staying there for a time $T_{ser}$ before recovering ($R$). On the other 
hand, infected individuals who have not been tested are sent to the class $R$ after the 
infection time $T_{inf}$. Matter of fact, since no vaccine has been developed until today, 
the model does not contain an immunization term. No rigid quarantine is taken into account. 
Furthermore, in Eq.~(\ref{quarentena}), the factor $T_{ser}$ dividing $Q$ represents a rate 
of exit from the quarantine (to the group $R$).

%%%%%%%%%%%%%%%%%%%%%%%%%%%%%%%%%%%%%%%%%%%%%%%%%%%%%%%%%%%%%%%
\subsection{Adjusting parameters to real data}
\label{ts}
%%%%%%%%%%%%%%%%%%

It is known that for systems composed of differential equations with $r$ unknown 
parameters, $2r + 1$ experiments with real data are needed to obtain all the information 
that is potentially available about the parameters \cite{son02}. Since in our case we 
have only two adjustable parameters ($r=2$), we need at least $5$ real data to adjust 
parameters correctly. This minimum value is automatically taken into account in all 
numerical simulations when adjusting the parameters.

\begin{figure}[h]
	\centering
	\includegraphics[width=0.88\columnwidth]{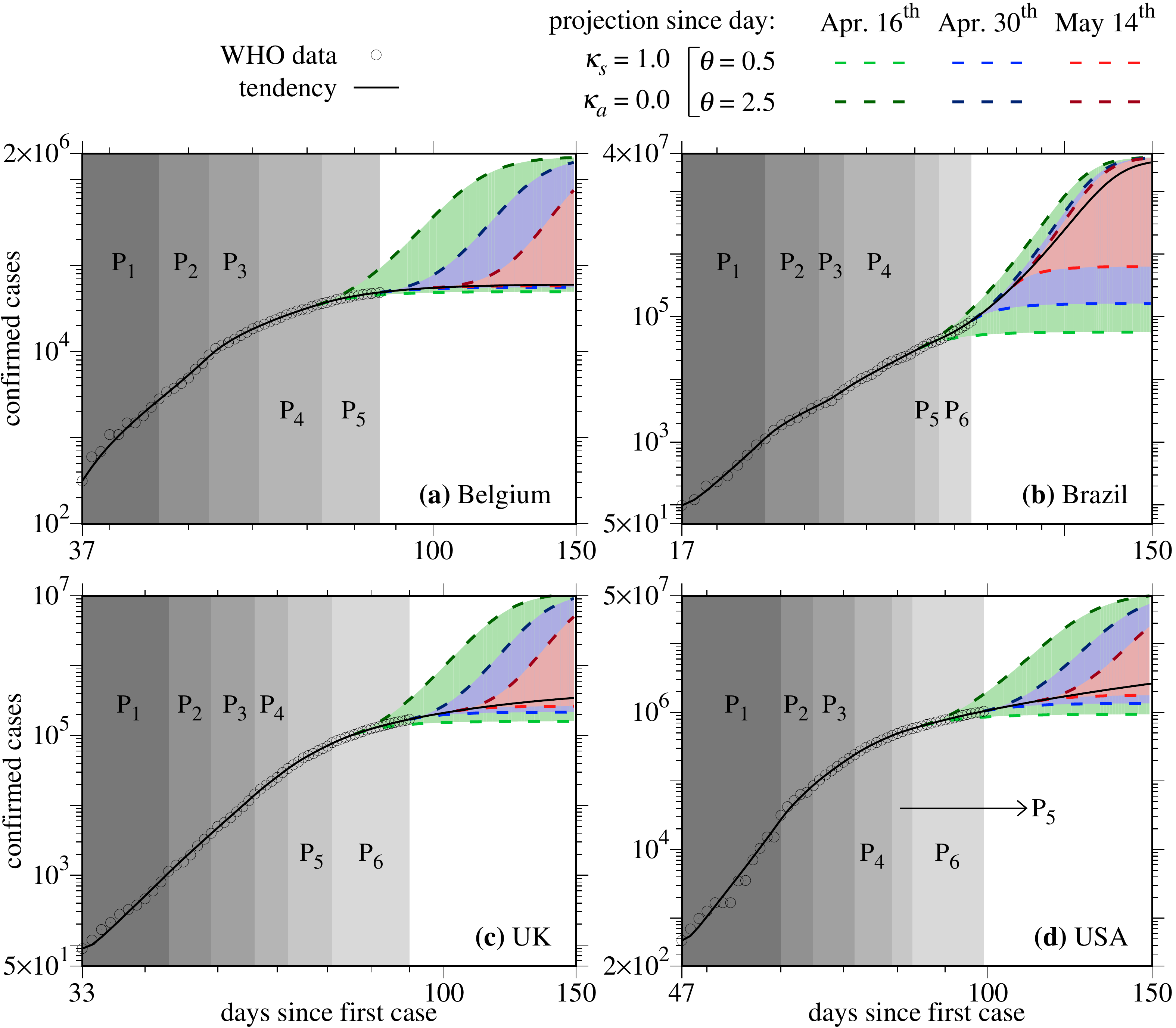}
	\caption{The cumulative number of confirmed cases of COVID-19 in the countries: (a) 
    Belgium, (b) Brazil, (c) UK, and (d) USA. Empty circles are the real data and the 
    black-continuous curves are the results from the simulations. Colors are related to 
    social distance actions applied in distinct days (discussed in the text).}
	\label{cenarios}
\end{figure}

Empty circles in Fig.~\ref{cenarios} are the real data for the cumulative number of 
confirmed cases of COVID-19 for the four countries analyzed. To find the best values for 
the pairs $(\theta,\kappa_s)=(\theta^{eff}, \kappa_s^{eff})$ that fit the real data and 
the best time-series split in periods P$_i$, we performed simulations varying $\theta \in 
[0.5,5.0]$ using a step equal to $0.1$ and $\kappa_s \in [0.0,1.0]$ using a step equal to 
$0.05$ and testing different combinations of periods P$_i$, always obeying the minimum 
amount of real data requested in each period. The goal of these simulations is to minimize 
the mean square error between the numerical results and real data. Thereon, we need five 
pairs of parameters in Fig.~\ref{cenarios}(a), namely P$_1$, P$_2$, P$_3$, P$_4$, and 
P$_5$ for Belgium, and six pairs of parameters for the other countries, seen in 
Fig.~\ref{cenarios},(b)-\ref{cenarios}(d). Details of the adjustable parameters are given 
in Table \ref{T1}. The initial condition $E(t_0)$ for the variable $E(t)$ is determined 
inside the first period P$_1$ of the data considering the interval $E(t_0) \in 
[C(t_0)/5,10C(t_0)]$ using a step equal $C(t_0)/5$, where $C(t_0)$ is the cumulative 
number of confirmed cases obtained from the WHO data for the first day in P$_1$. We do not 
start the parameter adjustment from the first day of reported infections, but later on. 
The model produces better results in such cases. 

After adjusting the parameters to the real data, the black-continuous curves in 
Fig.~\ref{cenarios} display the results of integrating equations of the model. We observe 
that these curves nicely reproduce the data in all cases. When real data are not available 
anymore, the black-continuous curves represent projections of the cumulative number of 
infected individuals until the day $150$, considering that the pair $(\theta^{eff}, 
\kappa^{eff}_s)$ found in the last period will not be changed. 

%%%%%%%%%%%%%%%%%%%%%%%%%%%%%%%%%%%%%%%%%%%%%
\begin{table}[!h]
  \caption{Values of $\theta^{eff}$ and $\kappa_s^{eff}$ for each period obtained by adjusting 
	  	the model to the real data. The last column shows the values of $C$ after 150 days 
	  	since the first case. }
	\centering
	\resizebox{\textwidth}{!}{
	\begin{tabular}{|l||c|c||c|c||c|c||c|c||c|c||c|c||c|}
		\hline
		\multicolumn{1}{|c||}{}                          &
        \multicolumn{2}{c||}{P$_1$}           & \multicolumn{2}{c||}{P$_2$}        & \multicolumn{2}{c||}{P$_3$}           & \multicolumn{2}{c||}{P$_4$}        & \multicolumn{2}{c||}{P$_5$}           & \multicolumn{2}{c||}{P$_6$}                     &                       \\ \cline{2-13}
		\multicolumn{1}{|c||}{\multirow{-2}{*}{{\bf Country}}} & $\theta ^{eff}$ & $\kappa _a ^{eff}$ & $\theta ^{eff}$ & $\kappa _a ^{eff}$ & $\theta ^{eff}$ & $\kappa _a ^{eff}$ & $\theta ^{eff}$ & $\kappa _a ^{eff}$ & $\theta ^{eff}$ & $\kappa _a ^{eff}$ & $\theta ^{eff}$      & $\kappa _a ^{eff}$      & \multirow{-2}{*}{$C$} \\ \hline
		Belgium                                         & $2.50$          & $1.00$             & $2.60$          & $0.00$             & $1.00$          & $0.00$             & $1.20$          & $0.80$             & $0.70$          & $0.40$             & \multicolumn{2}{c||}{\cellcolor[HTML]{C0C0C0}-} & $6.0\times 10^4$      \\ \hline
		Brazil                                          & $4.40$          & $0.20$             & $1.50$          & $0.20$             & $2.70$          & $0.00$             & $1.70$          & $0.50$             & $1.40$          & $1.00$             & $2.10$               & $0.45$                  & $2.9\times 10^7$      \\ \hline
		UK                                              & $3.60$          & $0.10$             & $2.80$          & $0.00$             & $2.60$          & $0.10$             & $1.90$          & $0.00$             & $1.30$          & $0.05$             & $1.00$               & $0.30$                  & $3.4\times 10^5$      \\ \hline
		USA                                             & $4.60$          & $0.05$             & $2.60$          & $1.00$             & $2.00$          & $0.40$             & $1.30$          & $0.75$             & $0.90$          & $1.00$             & $1.10$               & $0.15$                  & $2.6\times 10^6$      \\ \hline
	\end{tabular}}
	\label{T1}
\end{table}

%\section{Containment actions: past, present and future}
\section{Containment measures: early, current and late actions}
\label{ppf}

In this section, we discuss the effects of distinct strategies applied in different days 
on the total number of infected individuals. Essentially we discuss two strategies: 
\begin{description}
  \item[(i)] vary the degree of the social distance; 
  \item[(ii)] for a constant value of the social distance, vary the number of daily tests 
that allow identifying and isolating the infected asymptomatic individuals. 
\end{description} 

\subsection{The social distance effect}

As mentioned before, Fig.~\ref{cenarios} displays the real data (empty circles) and 
black-continuous curves which were adjusted to fit the data. The results are shown for 
Belgium in Fig.~\ref{cenarios}(a), Brazil in Fig.~\ref{cenarios}(b), UK in 
Fig.~\ref{cenarios}(c), and USA in Fig.~\ref{cenarios}(d). During the integration of the 
ODEs of the model, it is possible to change the parameter $\theta$ which represents the 
amount of social distance. Therefore, we changed this parameter from $0.5$ to $2.5$ using 
a step of $0.002$ in three distinct dates, namely April 16$^{\text{th}}$ (green-dashed 
curves), April 30$^{\text{th}}$ (blue-dashed curves), and May 14$^{\text{th}}$ (red-dashed 
curves). Curves with dark colors are related to $\theta=2.5$, and light colors to 
$\theta=0.5$. Worth mention that high values of $\theta$ mean low degrees of social 
distance, what can potentialize the epidemic spread. We clearly see, in the case of 
Belgium for example, that strong social distance strategies ($\theta=0.5$) can flatten the 
curves for the three distinct days. However, their efficiency in flattening the curves 
becomes less for later days (see blue and red light dashed curves). On the other hand, if 
social distance strategies are relaxed to $\theta=2.5$ in Belgium, when compared to 
$\theta^{eff}=0.7$ obtained for the last period P$_5$ (see Table \ref{T1}), the number of 
infected people increases very much. Furthermore, if you wait longer to relax the social 
distance, days April 30$^{\text{th}}$ (blue-dashed curves) or May 14$^{\text{th}}$ 
(red-dashed curves), for example, the total number of infected cases diminishes. 
Essentially the same behavior is observed for all the other countries analyzed. Please see 
Figs.~\ref{cenarios}(b)-\ref{cenarios}(d). In all these simulations the values 
$\kappa_s=1.0$ and $\kappa_a=0$ were kept fixed.

\subsection{Testing asymptomatic individuals}

\begin{figure}[!t]
	\centering
	\includegraphics[width=0.88\columnwidth]{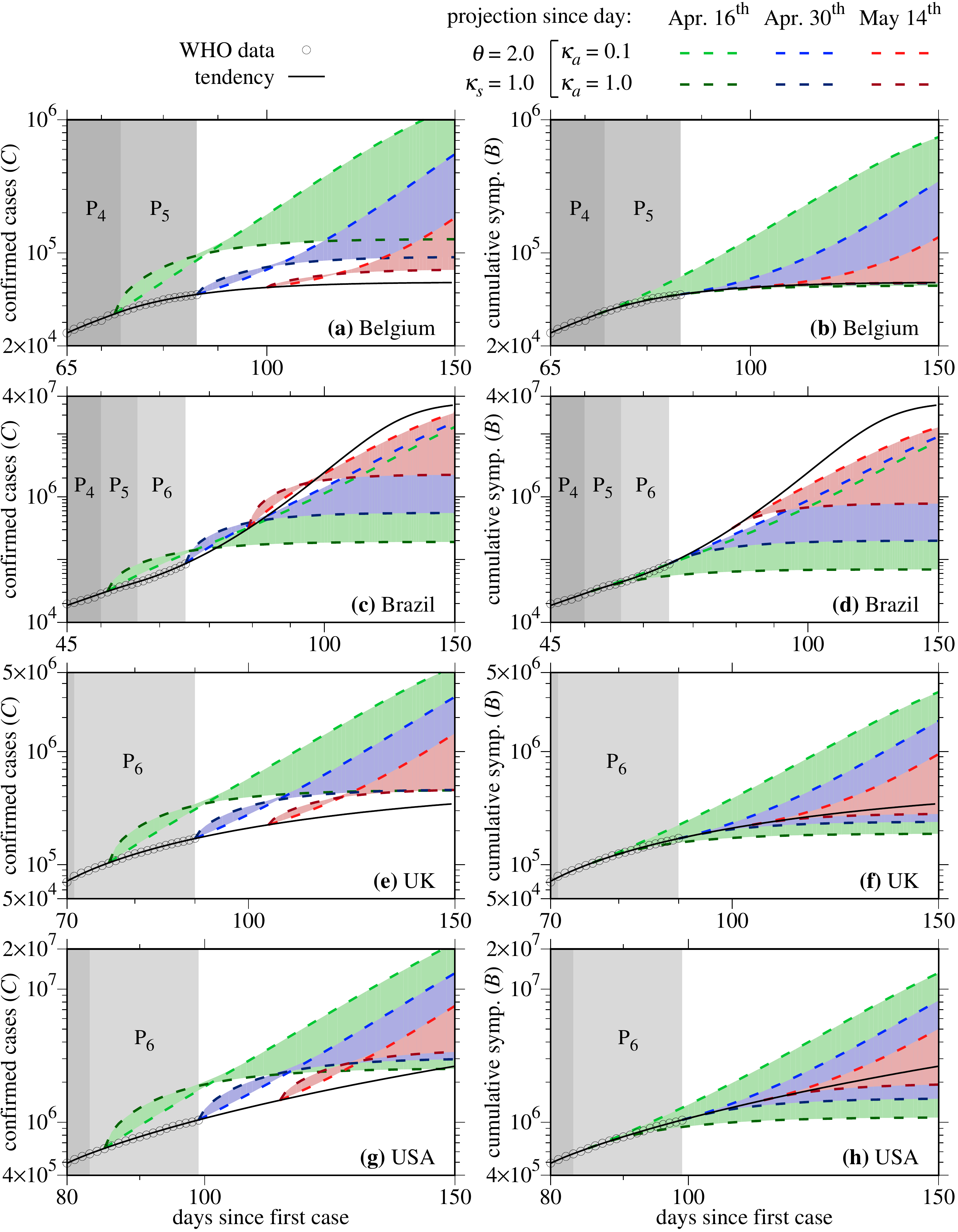}
	\caption{The left column displays the total cumulative number of confirmed cases of 
    COVID-19 for the same countries from Fig.~\ref{cenarios}, and the right column 
    displays the cumulative number of symptomatic infected individuals for the same 
    countries. Colors are related to the starting of the strategy of testing 
    asymptomatic infected individuals in distinct days (discussed in the text).} 
	\label{cenarios2}
\end{figure}

At next, we keep the social distance parameter constant at $\theta=2.0$, set $\kappa_s=1.0$, 
and change the daily rate of identification of asymptomatic infected individuals. The 
choice for this strategy is that without tests it is {\it impossible} to recognize that 
asymptomatic individuals are infected. In the simulations, we varied $\kappa_a$ from $0.1$ 
to $1.0$ using a step of $0.009$. Results are presented in Fig.~\ref{cenarios2} for the 
same countries from Fig.~\ref{cenarios}. For better visualization, we start the plot at 
later times when compared to Fig.~\ref{cenarios}. Figures \ref{cenarios2}(a), 
\ref{cenarios2}(c), \ref{cenarios2}(e), and \ref{cenarios2}(g) display the total 
cumulative number of confirmed infected cases and Figs.~\ref{cenarios2}(b), 
\ref{cenarios2}(d), \ref{cenarios2}(f), and \ref{cenarios2}(h) show the cumulative number 
of only symptomatic infected individuals. Strategies are again applied in days April 
16$^{\text{th}}$ (green-dashed curves), April 30$^{\text{th}}$ (blue-dashed curves), 
and May 14$^{\text{th}}$ (red-dashed curves). Curves with dark colors are related to 
$\kappa_a=1.0$ and curves with light colors to $\kappa_a=0.1$. This constant can be 
interpreted as follows: $\kappa_a=0.1$, for example, represents a daily rate of 
identification and isolation of $10\%$ of all asymptomatic infected individuals. This 
represents a huge number of daily tests for countries like Brazil and USA, which have a 
large population.

To make it possible to compare the projection tendencies, for which $\kappa_a = 0$, and 
the scenarios shown in Fig.~\ref{cenarios2}, we compute the cumulative number of 
symptomatic infectious cases ($B$). In this quantity, asymptomatic cases or those with 
mild symptoms are not considered. Similar to the $C$ variable, $B$ is an auxiliary 
variable of the model, obtained by integrating the ODE
\begin{align}
	\frac{dB}{dt} = (1 - \beta)\frac{E}{T_{lat}}.
\end{align}

Let us discuss results from Fig.~\ref{cenarios2} taking just one country: UK. In 
Fig.~\ref{cenarios2}(e) we observe that the strategy of realizing tests on asymptomatic 
individuals instantly increases the total number of confirmed cases. For the largest 
value $\kappa_a=1.0$ for example, the dark green curve increases very much on the day 
April 16$^{\text{th}}$. For $\kappa_a=0.1$, the light green curve barely changes in this 
day. However, after around $15$ days, both curves cross each other and the dark green 
curve asymptotically converges to a much smaller value than the light green curve, which 
shows that the realization of a huge amount of daily tests to identify and isolate 
asymptomatic individuals is also an efficient strategy that could be applied to relax the 
social distance (increase the value of $\theta$). Nevertheless, it is important to mention 
that, for countries with large populations, values $\kappa_a \approx 1.0$ are not 
practical parameters. The same behavior can be observed when the tests are applied in days 
April 30$^{\text{th}}$ (blue-dashed curves) and May 14$^{\text{th}}$ (red-dashed curves). 
Essentially an analogous interpretation is valid for the other countries. One difference 
is observed for Brazil. In Fig. \ref{cenarios2}(c) we can see that all the dashed curves 
cross the black-continuous curve of the tendency, meaning that even the late actions were 
able to diminish the number of infected individuals. It occurs because the constant value 
$\theta=2.0$ is lower than the $\theta^{eff}=2.1$ obtained in the last period P$_6$ for 
Brazil (see Table \ref{T1}). For the other countries, this is not the case since the 
black-continuous curves had a smaller value of $\theta^{eff}$ at P$_6$, or P$_5$ for 
Belgium (see Table \ref{T1}), when compared to $\theta=2.0$ used in Fig.~\ref{cenarios2}.  

At next, we discuss some projections for the cumulative number of symptomatic infected 
individuals, shown in Figs. \ref{cenarios2}(b) for Belgium, \ref{cenarios2}(d) for Brazil, 
\ref{cenarios2}(f) for UK, and \ref{cenarios2}(h) for USA. Now, we take the example of USA. 
When increasing the value of $\theta$ from $\theta^{eff}=1.10$ (see Table \ref{T1}) to 
$\theta=2.0$ and setting $\kappa_a=1.0$, on April 16$^{\text{th}}$, the dark green curve 
tends to flatten the growth of the cumulative number of symptomatic infected cases. On the 
other hand, for $\kappa_a=0.1$, the tendency is to increase such quantity when compared to 
the black-continuous curve. This projects bad news for USA in case they relax the social 
distance (increase the value of $\theta$) and apply a small number of tests to the 
identification of asymptomatic infected individuals. Similar behavior is observed for the 
other countries. The only difference is for Brazil, shown in Fig. \ref{cenarios2}(d), 
where the black-continuous curve is lying above the dashed curves once the value of 
$\theta$ used in these strategies is lower than the $\theta^{eff}$ obtained in P$_6$ for 
Brazil. The projections tend to get worst as the application day of the strategy is 
delayed.

%%%%%%%%%%%%%%%%%%%%%%%%%%%%%%%%%%%%%%%%%%%%%%%%%%%%%%%%%%%%%%%%%%%%%%%%%
\section{Conclusions}
\label{conclusions}
%%%%%%%%%%%%%%%%%%%%%%%%%%%%%%%%%%%%%%%%%%%%%%%%%%%%%%%%%%%%%%%%%%%%%%%%%

The cumulative number of confirmed cases of COVID-19 until April 30$^{\text{th}}$, 
2020, is demonstrated for four exemplary countries: Belgium, Brazil, UK, and USA, 
representing three distinct continents. After an initial period with a low incidence of 
newly infected people, a power-law growth of the number of confirmed cases is observed. 
For each country, we found a distinct growth exponent. USA leads the increasing rate, 
followed by UK, Brazil, and Belgium. For Belgium, UK, and USA, countries with a large 
number of infected individuals, the power-law growth gave place to a distinct behavior 
when approaching saturation. Brazil is still in the power-law regime. Such updates of the 
data and projections corroborate recent results regarding the power-law growth of the 
cumulative number of infected individuals by the new coronavirus and its strong 
correlation between different countries around the world \cite{elb20-1}.

Furthermore, we study a variation of the well known SEIR epidemic model 
\cite{modeloSeir2,modeloSeir1} for predictions using (or not) distinct government 
strategies applied on three distinct dates, namely April 16$^{\text{th}}$ (early action), 
April 30$^{\text{th}}$ (current action), and May 14$^{\text{th}}$ (late action). The main 
goal is to show that time is one of the most relevant weapons we have in the battle 
against the COVID-19. It has been shown recently that there is a short time window for 
which it is possible to avoid the spread of the epidemic \cite{gio20}. In our case, in 
the three days mentioned above, we applied two strategies: (i) distinct degrees of social 
distance (vary $\theta$), and (ii) distinct degrees of identification of asymptomatic 
individuals (vary $\kappa_a$). In the first strategy, we change the values of $\theta$ 
from $0.5$ to $2.5$, meaning strong and essentially no social distance containments, 
respectively. In the second strategy, we change $\kappa_a$ from $0.1$ to $1.0$. This can 
be interpreted as identifying daily $10\%$ of all asymptomatic individuals when 
$\kappa_a=0.1$. The ideal case is represented using $\kappa_a=1.0$, when all asymptomatic 
infected people are identified each day. Results for all countries convince us that 
non-pharmacological strategies must be applied as soon as possible. These include social 
distance and a large number of testing and immediate isolation of asymptomatic infected 
individuals. Furthermore, time delays in applying such strategies lead to an irreversible 
catastrophic number of infected people.

%%%%%%%%%%%%%%%%%%%%%%%%%%%%%%%%%%%%%%%%%%%%%%%%%%%%%%%%%%%%%%%
\section*{Acknowledgments}
%%%%%%%%%%%%%%%%%%%%%%%%%%%%%%%%%%%%%%%%%%%%%%%%%%%%%%%%%%%%%%%
The authors thank CNPq (Brazil) for financial support (grant numbers 
432029/2016-8, 304918/2017-2, 310792/2018-5 and 424803/2018-6), and they also 
acknowledge computational support from Prof.~C. M. de Carvalho at LFTC-DFis-UFPR 
(Brazil). C. M. also thanks FAPESC (Brazilian agency) for financial support.

%%%%%%%%%%%%%%%%%%%%%%%%%%%%%%%%%%%%%%%%%%%%%%%%%%%%%%%%%%%%%%%
%\section*{References}
%%%%%%%%%%%%%%%%%%%%%%%%%%%%%%%%%%%%%%%%%%%%%%%%%%%%%%%%%%%%%%%

%\bibliography{references}

\end{document}